\documentclass[12pt,amsmath,amssymb]{article}

\newcommand{\be}{\begin{equation}}
\newcommand{\ee}{\end{equation}}
\newcommand{\bea}{\begin{eqnarray}}
\newcommand{\eea}{\end{eqnarray}}

\begin{document}
\begin{titlepage}

\begin{flushright}
KEK-TH-1243\\
\today
\end{flushright}

\vspace{1in}

\begin{center}

{\bf Compactification of IIB Theory with Fluxes and Axion-Dilaton String Cosmology}

\vspace{1in}

\normalsize

{  Eiji Konishi$^{1}$\footnote{E-mail address: konishi.eiji@s04.mbox.media.kyoto-u.ac.jp} and 
Jnanadeva Maharana$^{2,3}$\footnote{E-mail address: maharana@iopb.res.in}}

\normalsize
\vspace{.5in}

 {\em $^1$Faculty of Science, Kyoto University, Kyoto 606-8502, Japan\\
$^{2}$National Laboratory for High Energy Physics (KEK), Tsukuba, Ibaraki 305, 
Japan\\ 
$^{3}$Institute of Physics, Bhubaneswar-751005, India\footnote[4]{Present and permanent address}}

\end{center}

\vspace{1in}

\baselineskip=24pt
\begin{abstract}
Compactification of type IIB theory on torus, in the presence of fluxes, is considered. The reduced effective action is expressed in manifestly S-duality invariant form. Cosmological solutions of the model are discussed in several cases in the Pre-Big Bang scenario.

\end{abstract}

\vspace{.7in}
 
\end{titlepage}

\section{Introduction}
It is recognized that string theory offers the prospect of unifying the
fundamental forces of Nature \cite{book}. The developments in string theory 
have shed lights in our understanding of the physics of black holes and have
addressed important problems in cosmology. Furthermore, there is a lot of
progress to establish connections between string theory and the standard
model of particle physics which comprehensively explains a vast amount of
experimental data.  
String theory is endowed with a rich symmetry structure. Notable among them are
dualities \cite{r1,r2,r3}. 
The strong-weak duality, S-duality, relates strong and weak coupling
 phases. In some cases these two phases of the same theory may be related and 
in some other cases strong and weak coupling regimes of two different theories 
are S-dual to each other. Type IIB string theory is an example of the former 
whereas, to recall a familiar example,
 heterotic string with $SO(32)$ gauge group is related by S-duality to 
type I theory with same gauge group in $D=10$. The T-duality, which we mention 
in passing, is tested perturbatively. A simple illustration is to consider 
compactification of a spatial coordinate of a bosonic closed string 
on $S^1$ of radius $R$. The 
perturbative spectrum of this theory matches with the one where the 
corresponding spatial coordinate is compactified on a circle of reciprocal 
radius. It is worthwhile to note that the web of dualities have provided us 
powerful tools to understand string dynamics in diverse dimensions.

The purpose of this article is to envisage toroidal compcatification of type 
IIB string theory in presence of constant 3-form fluxes along compact 
direction. 
Our goal is to investigate consequences of S-duality for reduced string 
effective action and examine possible mechanism for breaking of S-duality. Moreover, we elaborate a few points and sharpen them which were not addressed in the 
present perspectives in earlier works in the context of a four dimensional effective action where the symmetry is enhanced. 

The vanishingly small value of cosmological constant, $\Lambda_c$, might be 
understood by invoking the naturalness argument as has been advocated by  
us \cite{m1,m2,jim,k1}.
Although there are several proposals to explain smallness of 
the cosmological constant, 
this issue has not been completely resolved to every one's satisfaction.
It is to be noted that at the present juncture S-duality does not manifest as 
an exact symmetry of Nature. In other words, one of the consequences of S-duality would be to discover magnetically charged particles and dyons. In this respect, we may mention that we have no conclusive experimental evidence so far in favor of S-duality as a symmetry in the domain of low energy. Moreover, we have not found evidence for massless 
dilaton and axion from on going experiments. The axion, one often discusses in string theory framework, finds itself in a difficult position to be identified with the axion that one is 
looking for in various experiments 
which appears in the standard model phenomenology and GUT phenomenology \cite{ed,kim}.
Furthermore, it may be argued from the cosmological considerations that the dilaton should acquire its vacuum expectation value rather early in the evolution of the Universe (see section 2).
Thus the dilaton and axion are expected to acquire mass, and S-duality may be a 
broken symmetry of Nature. As we shall see, for the model at hand, the reduced 
action does not admit a cosmological constant term when S-duality is preserved.
 If S-duality is spontaneously broken nonzero (positive) cosmological 
constant appears in our model. It is of interest to explore how S-duality is 
broken. 
Recently, spontaneous breaking of S-duality has been proposed by one 
of us by invoking the idea of gauging the $SL(2,{\bf{R}})$ group \cite{k2}. 
Our proposal is, if
 we invoke naturalness argument due to 't Hooft, the cosmological constant 
should 
remain small in the theory we are dealing with.
We are aware that the scenario we have envisaged is not close to the Universe 
described by the standard model. However, we have provided a concrete example 
realizing the arguments of naturalness to qualitatively argue  why $\Lambda_c$ 
could be small. We are also aware that there are subtle issues related with 
non-compact symmetries and their breaking. We have not addressed to those 
problems here.

In recent years, compactification of string theories with fluxes has attracted 
considerable attention 
\cite{f1,f2,f3,frev1,frev2,frev3}. In the context of brane world 
scenario such compactifications have assumed special significance. There is a 
lot of optimism that this approach might eventually allow us to build model 
which will realize the well tested standard model of particle physics although 
we have not achieved this objective completely. There has been a 
lot of progress in the string landscape approach to study the rich vacuum
structure of string theory and construct models which are close to 
the phenomenological descriptions of particle physics. 
It has been shown that the
landscape scenario admits a de Sitter phase with appropriate background
configurations and a novel mechanism to break supersymmetry\cite{kklt}. 
The toroidal compactification and the symmetries of reduced effective action 
have been studied extensively \cite{r1,r2,r3,ss,ms,hs}. 
Recently, compactification of heterotic string 
effective action on $d$-dimensional torus with constant fluxes along compact 
direction has been investigated in detail \cite{km1}.
 It was shown that the reduced action
 (with constant fluxes) can be cast in $O(d,d)$ invariant form. Moreover, the 
moduli acquire potentials which would otherwise not appear in standard 
dimensional reduction of the 10-dimensional action of heterotic theory.

As we shall show in sequel, the toroidal compactification of type IIB theory in
 the presence of fluxes associated with NS-NS and R-R backgrounds can be 
expressed in an S-duality invariant form. The type IIB action can be
 compactified on a torus to obtain manifestly S-duality invariant reduced
effective action following the standard procedure \cite{jm1}. In the present 
context, due to the coupling of axion-dilaton ${\cal{M}}$-matrix to the fluxes
 (see later) a potential term appears.  
We note 
that when we toroidally compactify the action and allow constant fluxes along 
the compact direction the resulting reduced action is no longer invariant under
 supersymmetry. The supersymmetry can be restored by adding appropriate 
sources \cite{sources}.  
However, our purpose is to investigate S-duality attributes of the reduced 
action. Moreover, as we shall elaborate in the following section, we argue that 
there might be a symmetry consideration in order to understand smallness of the
 cosmological constant.
Furthermore, when we focus our attention on $D=4$, we can dualize the 
space-time dependent 3-form field strength and express the effective action in 
yet another S-duality invariant form as has been noted earlier \cite{clw}.

We study the cosmological solutions of the effective action \cite{cr,gvr}.
Since in the present scenario axion-dilaton potential emerges due to the coupling of the doublet to the 
fluxes, it provides an opportunity to reexamine some of the well known 
issues in string  cosmology.
Notice that the fluxes dictate the form of the potential. 
Thus, unlike in the past, we have a frame work to introduce a potential
in the effective action whose structure follows from the symmetry
considerations. We study the 
graceful exit problem that arises in the context of Pre-Big Bang cosmology 
\cite{gvr,gv1}. 
In a simple case, we truncate the action by setting some of the scalars to zero
(although we retain the axion and dilaton); however the axion-dilaton potential
is kept in tact while  addressing the graceful exit issue \cite{bv}. 
We find that the no-go theorem of Kaloper,
 Madden and Olive is still valid for our model \cite{kmo}. 
We consider another form of truncated 
action to examine whether the Pre-Big Bang solution accompanied by scale factor
 duality \cite{v1} is admissible. Indeed, for this case we obtain a 
solution which  satisfies the scale factor duality. However, the dilaton blows 
up at an instant when scale  factor approaches a certain value \cite{gvr}. 
Therefore, there is an epoch where the tree 
level string effective action is not trustworthy and the loop corrections have 
to be accounted for. 

The paper is organized as follows. The next section is devoted to dimensional 
reduction of the ten dimensional action when fluxes are present along compact directions. It is shown that, for a simple compactification scheme, when $D=4$, the action can be expressed in manifestly $SL(3,{\bf{R}})$ invariant form where
 moduli parameterize the coset $\frac{SL(3,{\bf{R}})}{SO(3)}$, although this 
result was known for a while \cite{clw, jm3}; 
we discuss this aspect from another point of view.
 Section III is devoted to solving the equations of motion. We consider 
different (truncated) version of the 4-dimensional action to obtain 
cosmological solutions.
It is worthwhile to point out that during the initial developments of string 
cosmology the potentials were introduced by hand to explore various  
cosmological solutions. One of our motivations to study string cosmology
starting from a nonsupersymmetric model is that it 
is expected that in the early Universe supersymmetry might not be preserved.
The fourth section is devoted to discussions about the no-go theorem. A short 
appendix contains a detail calculation of axion-dilaton and scale factor evolution in the context of graceful exit problem.

\section{Effective Action}
The massless excitations of type IIB string theory consist of dilaton, $\hat{\phi}$, axion,
$\hat{\chi}$, graviton, $\hat{g}_{MN}$, two 2-form potential,
$\hat{B}_{MN}^{(i)}$, ($i=1,2$) and a 4-form potential $\hat{C}_{MNPQ}$ with self-dual field strength. Note however, that a covariant 10-dimensional effective action for type IIB theory cannot be written down when we want to incorporate 5-form 
self-dual field strength. In what follows, we present the 10-dimensional action
 in the string frame metric and do not include contribution of the field strength of $\hat{C}_{MNPQ}$. This omission does not affect our study of the symmetry properties of the action. As a notational convention we denote the fields in
 10-dimensions with a hat. The action is 
\begin{eqnarray}
&&
\hat{S}=\frac{1}{2}\int d^{10}x\sqrt{-\hat{G}}\biggl\{e^{-2\hat{\phi}}
\biggl(\hat{{{R}}}-\frac{1}{12}\hat{{{H}}}_{MNP}^{(1)}\hat{{{H}}}^{(1)\ MNP}
+4(\partial\hat{\phi})^2\biggr)
-\frac{1}{2}(\partial \hat{\chi})^2\nonumber\\&&-\frac{1}{12}\chi^2
\hat{{{H}}}^{(1)}_{MNP}\hat{{{H}}}^{(1)\ MNP}
-\frac{1}{6}\hat{\chi}\hat{{{H}}}_{MNP}^{(1)}\hat{{{H}}}^{(2)\ MNP}-
\frac{1}{12}\hat{{{H}}}^{(2)}_{MNP}\hat{{{H}}}^{(2)\ MNP}\biggr\}\label{eq:eq1}\;.
\end{eqnarray}
It is more convenient to consider string effective action which is expressed in
 terms of Einstein frame metric $\hat{g}_{MN}$ while discussing S-duality 
transformations and the invariance properties of the action. The two metrics 
are related by $\hat{g}_{MN}=e^{-\frac{1}{2}\hat{\phi}}{\hat G}_{MN}$, and the 
corresponding action is \cite{hull, jhs}
\begin{eqnarray}&&\hat{S}_E=\frac{1}{2\kappa^2}\int d^{10}x\sqrt{-\hat{g}}
\biggl\{\hat{{{R}}}_{\hat{g}}+
\frac{1}{4}{\mathrm{Tr}}\bigl(\partial_N\hat{{\cal{M}}}\partial^N\hat{{\cal{M}}}^{-1}\bigr)-
\frac{1}{12}\hat{{\bf{H}}}^T_{MNP}\hat{{\cal{M}}}\hat{{\bf{H}}}^{MNP}
\biggr\}\label{eq:action}\end{eqnarray}
where the axion-dilaton moduli matrix $\hat{{\cal{M}}}$ and the H-fields are 
defined as follows
\begin{equation}\hat{{\cal{M}}}=\left(\begin{array}{cc}\hat{\chi}^2
e^{\hat{\phi}}+e^{-\hat{\phi}}&\hat{\chi}e^{\hat{\phi}}\\\hat{\chi}
e^{\hat{\phi}}&e^{\hat{\phi}}\end{array}\right)\;,\ \ \ \hat{{\bf{H}}}_{MNP}
=\left(\begin{array}{c}\hat{{{H}}}^{(1)}\\\hat{{{H}}}^{(2)}\end{array}\right)_{MNP}\;.\end{equation}
The action Eq.(\ref{eq:action}) is invariant under the transformations
\begin{equation}\hat{{\cal{M}}}\to \Lambda\hat{{\cal{M}}}\Lambda^T\;,\ \ \ 
\hat{{{H}}}\to(\Lambda^T)^{-1}{{H}}\;,\ \ \ \hat{g}_{MN}\to\hat{g}_{MN}\;,
\end{equation}
 $\Lambda\in SL(2,{\bf{R}})$ and $\Sigma$ is metric of $SL(2,{\bf{R}})$ 
\begin{equation}\Sigma=\left(\begin{array}{cc}0&i\\-i&0\end{array}\right)\;,
\end{equation}
\begin{equation}\Lambda \Sigma\Lambda^T=\Sigma\;,\ \ \ 
\Sigma\Lambda\Sigma=\Lambda^{-1}\;,\end{equation}
\begin{equation}\hat{{\cal{M}}}\Sigma\hat{{\cal{M}}}=\Sigma\;,\ \ \ 
\Sigma\hat{{\cal{M}}}\Sigma=\hat{{\cal{M}}}^{-1}\;.\end{equation}
In order to facilitate toroidal compactification we choose the following upper 
triangular form for the vielbein
\begin{equation}\hat{e}_M^A=
\left(\begin{array}{cc}e_{\mu}^r&{\cal{A}}_\mu^\beta E^a_\beta\\0&E_\alpha^a
\end{array}\right)\;.\end{equation}
Here $M,N\cdots$ denote the global indices and $A,B\cdots$ the local Lorentz 
indices.
The 10-dimensional metric $\hat{g}_{MN}=e_M^Ae_N^B\eta_{AB}$, where $\eta_{AB}$
 is the 10-dimensional Lorentz metric. Here $\mu,\nu=0,1,2,\cdots,D-1$ and
$\alpha,\beta=D,\cdots,9$. Note that $r,s,\cdots$ denote the $D$-dimensional 
local indices whereas $a,b,\cdots$ are corresponding rest of the local indices,
 taking values $D,\cdots,9$.

Thus
\begin{equation}g_{\mu\nu}=e_\mu^r e_\nu^s \eta_{rs}\;,\ \ \ 
{\cal{G}}_{\alpha\beta}=E_\alpha^aE_\beta^b\delta_{ab}\;.\end{equation}
$\eta_{rs}$ is flat space Lorentzian metric.
With the above form of $E_M^A$ we note that
\begin{equation}\sqrt{-\hat{g}}=\sqrt{-g}\sqrt{{\cal{G}}}\;.\nonumber
\end{equation}
Let us denote the space-time coordinates as $\{x^\mu,\mu=0,1,\cdots,D-1\}$ and 
the compact coordinate of $T^d$ as $\{Y^\alpha,\alpha=D,\cdots,9\}$.

If we assume that the backgrounds do not depend on the compact coordinates 
$\{Y^\alpha\}$, then the 10-dimensional Einstein frame effective action reduces
 to \cite{ms,hs,jm1,ferr}
\begin{eqnarray}&&S_E=\frac{1}{2}\int d^Dx \sqrt{-g}\sqrt{{\cal{G}}}
\biggl\{R+\frac{1}{4}\bigl[\partial_\mu {\cal{G}}_{\alpha\beta}
\partial^\mu{\cal{G}}^{\alpha\beta}+\partial_\mu {\rm{ln}}{\cal{G}}
\partial^\mu {\rm{ln}}{\cal{G}}
-g^{\mu\lambda}g^{\nu\rho}{\cal{G}}_{\alpha\beta}
{\cal{F}}_{\mu\nu}^\alpha{\cal{F}}_{\lambda\rho}^\beta\bigr]\nonumber\\&&-
\frac{1}{4}{\cal{G}}^{\alpha\beta}{\cal{G}}^{\gamma\delta}\partial_\mu  
B_{\alpha\gamma}^{(i)}{\cal{M}}_{ij}\partial^\mu B_{\beta\delta}^{(j)}
-\frac{1}{4}{\cal{G}}^{\alpha\beta}g^{\mu\lambda}g^{\nu\rho}
H_{\mu\nu\alpha}^{(i)}{\cal{M}}_{ij}H_{\lambda \rho \beta}^{(j)}-
\frac{1}{12}H_{\mu\nu\rho}^{(i)}{\cal{M}}_{ij}H^{(j)\ \mu\nu\rho}\nonumber\\
&&+\frac{1}{4}{\rm{Tr}}(\partial_\mu{\cal{M}}\Sigma \partial^\mu{\cal{M}}
\Sigma) \biggr\}\label{eq:action7}\end{eqnarray}
by adopting the standard procedure for toroidal compactification of string 
effective action.

Note that the gauge fields ${\cal{A}}_\mu^\alpha$ appear due to the 
prescription of vielbein keeping in mind that the action will be dimensionally
reduced. When dimensional reduction is carried out these Abelian gauge fields
are associated with the $d$-isometries. 
 Moreover, 
the gauge fields $A_{\mu\alpha}^{(i)},i=1,2,\alpha=D,\cdots,9$ come from the 
dimensional reduction of $\hat{B}_{MN}^{(i)}$ as is well known. Besides scalars
 ${\cal{G}}_{\alpha\beta}$, additional set of scalars $B_{\alpha\beta}^{(i)}$ 
also appear from the reduction of $\hat{B}_{MN}^{(i)}$. 

The above action is expressed in the Einstein frame, ${\cal{G}}$ being 
determinant of ${\cal{G}}_{\alpha\beta}$. If we demand $SL(2,{\bf{R}})$ 
invariance of the above action, then the backgrounds are required to satisfy 
following transformation properties\footnote{If toroidal compactification of action eq.(\ref{eq:eq1}) is carried out then reduced action is not expressible in manifestly S-duality invariant form\cite{sroy}.}:

\begin{equation}
{\cal{M}}\to\Lambda{\cal{M}}\Lambda^T\;,\ \ \ 
H_{\mu\nu\rho}^{(i)}\to(\Lambda^T)^{-1}_{ij}H^{(j)}_{\mu\nu\rho}\;,
\end{equation}
\begin{equation}
A_{\mu\alpha}^{(i)}\to(\Lambda^T)^{-1}_{ij}A_{\mu\alpha}^{(j)}\;,\ \ \ 
B_{\alpha\beta}^{(i)}\to(\Lambda^T)^{-1}_{ij}B_{\alpha\beta}^{(j)}\;,
\end{equation}
and 
\begin{equation}g_{\mu\nu}\to g_{\mu\nu}\;,\ \ \ {\cal{A}}_\mu^\alpha\to 
{\cal{A}}_\mu^\alpha\;, \ \ \ {\cal{G}}_{\alpha\beta}\to{\cal{G}}_{\alpha\beta}\;.
\end{equation}

We focus our attention on the $D=4$ effective action and truncate it by setting
 some backgrounds to zero.
From now on we set ${\cal{A}}^\beta_\mu=0$. There are two 2-form fields 
$\hat{{{B}}}^{(i)}_{MN}$ coming from NS-NS and R-R sectors. When we compactify 
them to 4-dimensions, we keep only ${{B}}_{\mu\nu}^{(i)}$. The other field 
components ${{B}}_{\mu\alpha}^{(i)}=0$, and for the time being we also
set ${{B}}^{(i)}_{\alpha\beta}=0$.
The 4-dimensional action is
\begin{eqnarray}
&&S_E^{(4)}=\frac{1}{2}\int d^4x\sqrt{-G}\sqrt{{\cal{G}}}\biggl[{{R}}+
\frac{1}{4}\bigl\{\partial_\mu{\cal{G}}_{\alpha\beta}\partial^\mu
{\cal{G}}^{\alpha\beta}
+\partial_\mu{{\mathrm{ln}}}{\cal{G}}\partial^\mu {\mathrm{ln}}{\cal{G}}\bigr\}
-\frac{1}{12}{{H}}^{(i)}_{\mu\nu\rho}{\cal{M}}^{ij}{{H}}^{(j)\ \mu\nu\rho}
\nonumber\\&&+\frac{1}{4}{\mathrm{Tr}}(\partial_\mu{\cal{M}}\Sigma\partial^\mu
{\cal{M}}\Sigma)\biggr]\;.\end{eqnarray}

According to the compcatification procedure adopted above, the field strengths 
along compact directions are vanishing. The case of constant nonzero fluxes 
will be considered later.

We choose a simple compactification scheme where only a single modulus, $y(x)$,
 appears and 
we adopt the following form of the metric
\begin{equation}ds_{10}^2=g_{\mu\nu}dx^\mu dx^\nu+e^{y(x)/\sqrt{3}}dY^\alpha 
dY^\beta\delta_{\alpha\beta}\;.\label{eq:volume}\end{equation}

The resulting action is
\begin{eqnarray}S^{(4)}_E=\frac{1}{2}\int d^4x \sqrt{-g}e^{\sqrt{3}y}\biggl[{{R}}+
\frac{5}{2}\partial_\mu y\partial^\mu y -
\frac{1}{12}{{H}}_{\mu\nu\rho}^{(i)}{{\cal{M}}}_{ij}{{H}}^{(j)\mu\nu\rho}+
\frac{1}{4}{\mathrm{Tr}}(\partial_\mu{\cal{M}}\Sigma\partial^\mu 
{\cal{M}}\Sigma)\biggr]\;.\label{eq:action2}\end{eqnarray}

By rescaling the space-time metric, $g_{\mu\nu}$ we remove the 
over all factor of 
$e^{\sqrt{3}y}$ and bring the above action to the following form (we still denote the new space-time metric as $g_{\mu\nu}$).

The action is written by
\begin{eqnarray}S_E^{(4)}=\frac{1}{2}\int d^4x \sqrt{-g}\biggl[R-2(\nabla y)^2-
\frac{e^{2\sqrt{3}y}}{12}H_{\mu\nu\rho}^{(i)}{\cal{M}}_{ij}H^{(j)\mu\nu\rho}+
\frac{1}{4}{\rm{Tr}}(\partial_\mu {\cal{M}}\Sigma\partial^\mu{\cal{M}}\Sigma)
\biggr]\;.\label{eq:action5}\end{eqnarray}

Note that the 3-form field strengths $H_{\mu\nu\rho}^{(i)}$ can be dualized to 
trade for two pseudo scalars, $\sigma_i(x)$, $i=1,2$ in four dimensions. 
Moreover, the equations of motion for $H_{\mu\nu\rho}^{(i)}$ are conservation 
laws. Therefore, we expect that the five moduli 
$y(x)$, $\sigma_1(x)$, $\sigma_2(x)$, $\chi(x)$ and $\phi(x)$ will also reflect
 a symmetry. Indeed, they parameterize the coset 
$\frac{SL(3,{\bf{R}})}{SO(3)}$ \cite{clw}. 
Thus the action can be expressed in the 
following form
\begin{equation}S_E^{(4)}=\frac{1}{2}\int d^4x\sqrt{-g}\biggl[R+\frac{1}{4}{\rm{Tr}}
(\nabla U\nabla U^{-1})\biggr]\label{eq:action3}\end{equation}
using $SL(3,{\bf{R}})/SO(3)$ $U$-matrix
\begin{equation}U=e^{2\phi-\frac{2}{\sqrt{3}}y}
\left(\begin{array}{ccc}1&\chi&
\sigma_1-\chi\sigma_2\\\chi&\chi^2+e^{-2\phi}&\chi(\sigma_1-\chi\sigma_2)-
\sigma_2e^{-2\phi}\\\sigma_1-\chi\sigma_2&\chi(\sigma_1-\chi\sigma_2)-
\sigma_2e^{-2\phi}&(\sigma_1-\chi\sigma_2)^2+\sigma_2^2e^{-2\phi}+e^{-4\phi+2
\sqrt{3}y}\end{array}\right)\end{equation}
with $\det U=1$. 

Note that the equation of motion for the $U$-matrix is a conservation law, 

\begin{equation}\partial_\mu \bigl(\sqrt{-g}U^{-1}\partial^\mu U\bigr)=0
\end{equation}
representing five equations of motions since there are only five fields 
parameterizing $U$-matrix. When the action is expressed in terms of the scalars
 instead of the $U$-matrix, the corresponding equations of motion apparently 
give the impressions as if there are interaction potentials among the scalars.

Therefore, one is led to believe that there are $SL(2,{\bf{R}})$ invariant 
potential contradicting the result of \cite{jm1}. However, when suitable 
linear combinations of the equations of motion are taken, (derived from an 
action with component fields not of the form of the action (\ref{eq:action3})) indeed, one obtains five current conservation 
equations which match with the same number of conservation laws as one derives
from an $\frac{SL(3,{\bf{R}})}{SO(3)}$ symmetric action.

Thus far we have considered the scenario when backgrounds do not depend on 
compact coordinates. However, it is worthwhile to note that the effective 
action depends on the field strengths. Therefore, if constant field strengths 
are added to the effective action they will not affect the equations of motion 
modulo appearance of a cosmological constant term in certain cases. 
The appearance of cosmological constant term would be ruled 
out if additional symmetry constraints are imposed (for example, positive 
cosmological constant is not admissible if supersymmetry is desired).

Now let us examine how the contribution of the fluxes appears in the effective 
action. They contribute as the field strengths, 3-forms, coming from the NS-NS 
and R-R sectors. Thus their additional contribution to equation (\ref{eq:action2}) will be
\begin{eqnarray}
-\frac{1}{12}\sqrt{-g}\sqrt{{\cal{G}}}{{H}}_{\alpha\beta\gamma}^T
{\cal{M}}{{H}}_{\alpha^\prime\beta^\prime\gamma^\prime}
{\cal{G}}^{\alpha\alpha^\prime}{\cal{G}}^{\beta\beta^\prime}
{\cal{G}}^{\gamma\gamma^\prime}\;.
\end{eqnarray}

When we consider our simple compactification scheme, (after space-time metric has been rescaled to arrive at (\ref{eq:action5})) this term is
\begin{equation}-\frac{e^{-2\sqrt{3}y}}{12}\sqrt{-g}H_{\alpha\beta\gamma}^{(i)}
{\cal{M}}_{ij}H^{(j)}_{\alpha^\prime \beta^\prime\gamma^\prime}
\delta^{\alpha\alpha^\prime}\delta^{\beta\beta^\prime}
\delta^{\gamma\gamma^\prime}\end{equation}
with the appearance of this term the four dimensional effective action, 
$S_4^{D}$, (when $H_{\mu\nu\lambda}^{(i)}$ have been dualized) takes the form 
\begin{eqnarray}&&S_4^D=\frac{1}{2}\int d^4x \sqrt{-g}\biggl[R+\frac{1}{4}
{\rm{Tr}}(\partial_\mu{\cal{M}}^{-1}\partial^\mu{\cal{M}})-
2\partial_\mu y\partial^\mu y-\frac{e^{2\sqrt{3}y}}{2}\partial_\mu\sigma^T 
{\cal{M}}\partial^\mu\sigma\nonumber\\&& -\frac{e^{-2\sqrt{3}y}}{12}
{\cal{H}}_{\alpha\beta\gamma}^T{\cal{M}}{\cal{H}}^{\alpha\beta\gamma}\biggr]
\label{eq:SSB}\end{eqnarray}
where $\sigma=\left(\begin{array}{c}\sigma_1\\\sigma_2\end{array}\right)$ 
transforms as $SL(2,{\bf{R}})$ doublet like $\sigma\to(\Lambda^T)^{-1}\sigma$ 
and
${\cal{H}}_{\alpha\beta\gamma}=\left(\begin{array}{c}
{\cal{H}}_{\alpha\beta\gamma}^{(1)}\\
{\cal{H}}_{\alpha\beta\gamma}^{(2)}\end{array}\right)$
are the 3-form fluxes along compact directions.

The above action merits some discussions. We recall that in absence of 
${\cal{H}}_{\alpha\beta\gamma}$ the action (\ref{eq:action3}) is expressed in 
manifestly $SL(3,{\bf{R}})$ invariant form. The presence of the constant flux 
breaks $SL(3,{\bf{R}})$ symmetry. However, the above action is manifestly 
$SL(2,{\bf{R}})$ invariant. Moreover, notice the coupling of ${\cal{M}}$-matrix
 to the fluxes. Thus we have coupling of the $SL(2,{\bf{R}})$ multiplet to 
gravity and to a constant source term. It is well known that due to the 
coupling to the fluxes, the tadpoles make their appearances. Moreover, 
supersymmetry is not maintained in this scenario. As alluded earlier our goal 
is to explore various aspects of S-duality and we are aware that 
supersymmetries are not preserved. We would like to dwell on some other aspects
 due to the presence of fluxes in what follows.

The model under considerations is reminiscent of the works on chiral dynamics. 
It is worthwhile to note the issues addressed in the context of $\sigma$-model 
approach to pion physics. We might 
have a scenario where $SL(2,{\bf{R}})$ symmetry 
could be broken spontaneously. Alternatively, the S-duality could be broken 
explicitly if we assign specific configurations of the fluxes.
Although the last term in the above action is $SL(2,{\bf{R}})$ invariant, 
once we assign a specific configuration to the fluxes the symmetry is 
explicitly broken and still one has an axion-dilaton potential depending on the
choice of the flux that breaks the symmetry.  

We mention in passing that dilaton is expected to settle down to its ground state value 
in early epochs of the universe. It is well known that dilaton has two roles. 
It belongs to the massless spectrum of string theory. More importantly, the 
vacuum expectation value of dilaton determines gauge couplings, Yukawa 
couplings (hence fermionic masses) and a lot of other important parameters. To 
rephrase arguments to Damour and Polyakov \cite{dp}, 
dilaton must acquire its ground 
state value (say $\phi_0$) much before the era of nucleosynthesis. Otherwise, 
the so well tested results of big bang model in that era will be affected since
 delicate nuclear reactions depend crucially on masses of nucleons (and light
nuclei) and other fermions which in turn are 
determined by fermionic Yukawa couplings which can be computed in principle
from underlying string theory ( all these are controlled by vacuum expectation 
value of dilaton).

Furthermore, at the present epoch of the universe there is no conclusive 
experimental evidence for a weak (as weak as gravity) long range repulsive 
universal interaction.
Should dilaton remain massless, today, we expect such a repulsive force. 
Moreover, there are limits on dilaton mass to be $m_{{\mbox{dil}}}\ge 10^{-3}$
MeV.

If we consider the scenario of spontaneous symmetry breaking, we shall get a 
term like the cosmological constant as is obvious from eq.(\ref{eq:SSB}). 
Suppose the ${\cal{M}}$-matrix assumes a nonzero vacuum expectation value. 
Then, we may write ${\cal{M}}={\cal{M}}_0+\widetilde{{\cal{M}}}$, ${\cal{M}}_0$
 is the constant vacuum expectation value and $\widetilde{{\cal{M}}}$ is being 
the fluctuation, $<\widetilde{{\cal{M}}}>_0=0$.

Then
\begin{equation}
-\frac{1}{12}\sqrt{-g}{\cal{H}}_{\alpha\beta\gamma}^T{\cal{M}}_0
{\cal{H}}_{\alpha^\prime\beta^\prime\gamma^\prime}
{\delta}^{\alpha\alpha^\prime}{\delta}^{\beta\beta^\prime}{\delta}^
{\gamma\gamma^\prime}\end{equation}
is the cosmological constant term when the ten dimensional action is compactified on a torus of constant moduli, i.e. the radius of compactification is chosen to be a constant in eq. (\ref{eq:volume}) to start with (see discussion before eq.(\ref{eq:action4})).

\section{Equations of Motion} 
In this section we intend to present equations of motion associated with the actions 
considered in the previous section and look for solutions. We shall envisage 
cosmological scenario. It will be shown that we can obtain exact cosmological 
solutions for the Friedmann-Robertson-Walker (FRW) metric in certain cases. We note that some of the interesting aspects of type IIB string cosmology have been studied by several authors and a detail references can be found in the review of article of Copeland, Lidsey and Wands.\cite{clw} However, as mentioned above, in some of those investigations either axion-dilaton potential were not incorporated or the potentials were introduced on ad hoc basis. Our approach differs from earlier works in this sense. In what follows we 
shall consider the $k=0$ FRW metric, the spatially flat metric.

Let us first consider the equations of motion corresponding to action 
(\ref{eq:action3}).The matter field equation is that of the $SL(3,{\bf{R}})$ 
nonlinear $\sigma$-model coupled to gravity
\begin{equation}
\partial_\mu\bigl(\sqrt{-g}U^{-1}\partial^\mu U\bigr)=0\;.
\end{equation}
The Einstein equation is
\begin{equation}
R_{\mu\nu}-\frac{1}{2}g_{\mu\nu}R=-\frac{1}{8}\sqrt{-g}\partial_\mu U^{-1}
\partial_\nu U\;.
\end{equation}

In the cosmological scenario the FRW metric is
\begin{equation}
ds^2=-dt^2+a^2(t)dr^2+a^2(t)d\Omega^2\;.
\end{equation}
The equations of motion (using the 0-0 component of Einstein equation and 
matter equation)
\begin{equation}
16\dot{h}+24h^2-{\rm{Tr}}(\dot{U^{-1}}\dot{U})=0\;,
\end{equation}
with $h=\dot{a}/a$ as the Hubble parameter, and
\begin{equation}
\partial_t(a^3 U^{-1}\dot{U})=0\;.
\end{equation}
Thus
\begin{equation}
a^3U^{-1}\dot{U}=A\;,
\end{equation}
and $A$ is a constant $3\times 3$ matrix.

It follows from the Hamiltonian constraint that
\begin{equation}
6h^2+\frac{1}{4}{\rm{Tr}}(\dot{U^{-1}}\dot{U})=0\;.
\end{equation}

Thus combining these two equations, we solve for the scale factor and the 
$U$-matrix
\begin{equation}
a(t)\sim t^{1/3}\;,\ \ \ U=e^{A{\rm{ln}}t}\;.
\end{equation}
Note that in our approach we are able to present a cosmological solution for the FRW metric.  This is analogous to the solution obtained in \cite{mvc} while dealing with $O(d,d)$ symmetric action. Here we solve for the scale factor and the $U$-matrix 
parameterized in terms of the five fields, $\phi$, $\chi$, $y,$ $\sigma_1$ and 
$\sigma_2$. The solution involves the integration constant $A$ (now a 
$3\times3$ matrix with five independent components). We may recall here, 
in their approach to type IIB string cosmology, Copleland, Lidsey and Wands
\cite{clw}  
identified various $SL(2,{\bf{R}})$ subgroups of $SL(3,{\bf{R}})$ parameterized
 in terms of the different combinations of the above five fields besides the 
space-time metric. In that formulation, interactions among these fields 
(always involving derivatives or currents) were identified as potentials. Thus 
in the cosmological scenario, one encountered potentials. These authors found 
several types of solutions for homogeneous cosmologies for various 
$SL(2,{\bf{R}})$ subgroups of the duality group $SL(3,{\bf{R}})$. We would like to point out that from a purely group theoretic perspective, in the present approach, we are able to arrive the cosmological solution. 
This is an elegant and powerful method.

Let us turn our attention to the effective action $S_4^D$, eq. (\ref{eq:SSB}), 
where we have already dualized the three forms $H_{\mu\nu\rho}^{(i)}$ to 
$\sigma_i$, and we have taken into account the presence of fluxes. In the 
cosmological scenario under consideration, the equations of motion cannot be 
solved exactly. The presence of fluxes breaks $SL(3,{\bf{R}})$ to S-duality 
group $SL(2,{\bf{R}})$. If we examine for equations of motion, the variation of
 modulus, $y$, couples to ${\cal{M}}$-matrix and to 
$\sqrt{-g}\partial_\mu\sigma^T{\cal{M}}\partial^\mu \sigma$. The equations of 
motion for $\sigma_i$ are current conservations.

The following clarifying remark is in order at this stage.
Let us consider a case when the 10-dimensional theory has been compactified to 
the 4-dimensional theory with a constant radius of compactification. In other 
words, $y$, appearing in eq. (\ref{eq:volume}) carries no space-time 
dependence. In such a case, the scalar field $y(x)$ is constant and not 
dynamical to start with; however, $B_{\alpha\beta}^{(i)}(x)$ can still appear 
as (scalar) dynamical degree of freedom. This situation is different from the 
case where the volume modulus stabilizes to some constant value due to a 
mechanism built into the theory. Moreover, if one retains the modulus, $y(x)$, 
as dynamical in the reduced action and sets it to a constant value at the level
 of equation of motion, then the equation of motion for $y$ (although it is set
 to a constant at that stage) imposes a constraint equation for other fields. 
This situation is to be contrasted with the former situation where $y$ is 
already taken to be a constant right from the beginning when we compactify the 
10-dimensional action. To illustrate further, if we considered 
${\cal{G}}_{\alpha\beta}$ to be a constant then their corresponding kinetic 
energy term is eq.(\ref{eq:action7}) will be absent whereas all other terms 
will be present. In this paper when we use the phrase ``frozen modulus'' or 
freezing of modulus it is to be understood that $y(x)$ is set to be a constant 
when $D=10$ action is compactified. Let us consider the following simple form 
of the action where we set $\sigma_i=0$ and $y$ is a constant radius of 
compactification. The resulting action is
\begin{eqnarray}
S_4=\frac{1}{2}\int d^4x \sqrt{-g}\biggl[R+\frac{1}{4}{\rm{Tr}}(\partial_\mu 
{\cal{M}}\Sigma\partial^\mu {\cal{M}}\Sigma)-\frac{1}{12}
{\cal{H}}_{\alpha\beta\gamma}^T{\cal{M}}{\cal{H}}^{\alpha\beta\gamma}\biggr]\;.
\label{eq:action4}
\end{eqnarray}

We define the last term as $-{\rm{Tr}}SM$ to simplify the notations where

\begin{equation}
S=\frac{1}{12}\left(\begin{array}{cc}{\cal{H}}_{\alpha\beta\gamma}^{(1)}
{\cal{H}}^{(1)\alpha\beta\gamma}&{\cal{H}}_{\alpha\beta\gamma}^{(1)}
{\cal{H}}^{(2)\alpha\beta\gamma}\\{\cal{H}}_{\alpha\beta\gamma}^{(1)}
{\cal{H}}^{(2)\alpha\beta\gamma}&{\cal{H}}_{\alpha\beta\gamma}^{(2)}
{\cal{H}}^{(2)\alpha\beta\gamma}\end{array}\right)\;.
\end{equation}

The equation of motion associated with the ${\cal{M}}$-matrix is 
\begin{equation}
\partial_\mu \bigl(\sqrt{-g}\partial^\mu {\cal{M}}{\cal{M}}^{-1}\bigr)+
\sqrt{-g}{\cal{M}}S-\sqrt{-g}\Sigma S{\cal{M}}\Sigma=0
\end{equation}
in the matrix notations. The above equation is 
derived from the action (\ref{eq:action4}) keeping in mind that variation of 
${\cal{M}}$ is constrained since ${\cal{M}}\in SL(2,{\bf{R}})$. Furthermore, if
 ${\cal{M}}$ were an ordinary matrix (not constrained) then its variation of 
the last term in action (\ref{eq:action4}) will result in $S_{ij}$ in the 
equation of motion rather than combination of terms like ${\cal{M}}S$ and 
$S{\cal{M}}$. If we consider a situation where the $SL(2, {\bf R})$ symmetry
is broken spontaneously, then in this phase, we may express ${\cal M}=
{\cal M}_0+{\widetilde {\cal M}}$ where ${\cal M}_0$ is the constant $SL(2,{\bf R})$ matrix  and ${\widetilde {\cal M}}$ is its fluctuation.
Now let us turn our attention to 
Einstein equation. In spontaneously symmetry broken phase the constant flux 
does couple to both ${\cal M}_0$ and ${\widetilde {\cal M}}$. Notice that the
former is like a cosmological constant in the Einstein equation whereas
the latter contributes to the stress energy momentum tensor in the usual
fashion. 

Now let us discuss the cosmological scenario when the metric and all other 
fields depend only on the cosmic time. In the presence of the flux term, we shall 
get a complete set of equations if we include $y(t)$,  the modulus, and the 
other two sets of pseudo scalars, $\sigma_i(t)$, in the action. As noted 
earlier, the presence of flux introduces axion-dilaton potential, call it $\Lambda(\phi,\chi)$, which is $SL(2,{\bf{R}})$ invariant. We truncate the 
action with (and still the 
S-duality invariance is maintained) when moduli of torus is constant and 
$\sigma_i=0$. Notice that the presence of the S-duality invariant potential 
offers the prospect of exhibiting some of the interesting aspects of string 
cosmology.

We recall that, in the Pre-Big Bang proposal of Gasperini and Veneziano, one 
encountered the problem of graceful exit. It was observed that gravi-dilaton 
cosmology encountered the no-go theorem \cite{bv} 
such that classically the two regimes 
could not be smoothly connected under certain situations. 
It was hoped that axion-dilaton cosmology 
might overcome the difficulty. We shall examine the graceful exit issues in the following.

The 4-dimensional effective action now we consider (including the scalars
$B^{(i)}_{\alpha\beta}$) is
\begin{eqnarray}
&&S^{(4)}_E=\frac{1}{2}\int d^4x \sqrt{-g}\biggl[R+\frac{1}{4}{\rm{Tr}}(
\partial_\mu{\cal{M}}^{-1}\partial^\mu{\cal{M}})-
\frac{1}{4}\partial_\mu B_{\alpha\beta}^{(i)}{\cal{M}}_{ij}
\partial^\mu B^{(j)\alpha\beta}\nonumber\\&&-\frac{1}{2}\partial_\mu \sigma^T {\cal{M}}
\partial^\mu\sigma-\frac{1}{12}
{\cal{H}}_{\alpha\beta\gamma}^{(i)}
{\cal{M}}_{ij}{\cal{H}}^{(j)\alpha\beta\gamma} - 2(\partial y)^2\biggr]\;.
\end{eqnarray}
In a cosmological scenario where modulus is frozen, the above action goes
over to
\begin{eqnarray}
&&S_E ^{(4)}=\frac{1}{2}\int dt \sqrt{-g}\biggl[R-\frac{1}{4}{\rm{Tr}}
\dot{{\cal{M}}^{-1}}\dot{{\cal{M}}}+\frac{1}{4}\dot{B}_{\alpha\beta}^{(i)}
{\cal{M}}_{ij}\dot{B}^{(j)\alpha\beta}+\frac{1}{2}\dot{\sigma}^T{\cal{M}}
\dot{\sigma}\nonumber\\&&-\frac{1}{12}{\cal{H}}_{\alpha\beta\gamma}^{(i)}
{\cal{M}}_{ij}{\cal{H}}^{(j)\alpha\beta\gamma}\biggr]\;.\label{eq:action6}
\end{eqnarray}

Equations of motion are
\begin{eqnarray}
&&\frac{d}{dt}(\sqrt{-g}{\cal{M}}_{ij}\dot{B}^{(j)}_{\alpha\beta})=0\;,\label
{eq:EOM4}\\&&\frac{d}{dt}(\sqrt{-g}{\cal{M}}_{ij}\dot{\sigma}_j)=0\;,
\label{eq:EOM5}\\
&&\frac{d}{dt}(\sqrt{-g}(\dot{{\cal{M}}}{\cal{M}}^{-1}))+\sqrt{-g}
{\cal{M}}{\widetilde{S}}-\sqrt{-g}\Sigma {\widetilde{S}}{\cal{M}}\Sigma=0\label{eq:EOM6}
\end{eqnarray}
 where $\widetilde{S}$ also contains terms involving
\begin{equation}
{\widetilde{S}}=\frac{1}{12}{\cal{H}}_{\alpha\beta\gamma}^{(i)}
{\cal{H}}^{(j)\alpha\beta\gamma}-\frac{1}{2}\dot{\sigma}^i\dot{\sigma}^j-
\frac{1}{4}\dot{B}^{(i)}_{\alpha\beta}\dot{B}^{(j)\alpha\beta}\;.
\end{equation}

The first two equations (\ref{eq:EOM4}) and (\ref{eq:EOM5}) imply that the 
time integrations  are
\begin{equation}
{\cal{M}}_{ij}\dot{B}_{\alpha\beta}^{(j)}=C_{\alpha\beta}^i\;,\ \ \ \sqrt{-g}
{\cal{M}}_{ij}\dot{\sigma}_j=D_i\label{eq:EOM7}\end{equation}
 $C$ and $D$ being time independent.

Thus when we consider equations of motion (\ref{eq:EOM6}) associated with the 
${\cal M}$-matrix and utilize the relation (\ref{eq:EOM7}) in them in the definition of
${\tilde S}$, then we notice that elimination of the time derivatives of
$B^{(i)}_{\alpha\beta}$ and $\sigma ^i$ from $\tilde S$ leads to additional
potential terms that depend on ${\cal M}$-matrix.  

Let us discuss the presence of the axion-dilaton potential (the last term) in 
the action. One might hope that with the presence of a potential 
$\Lambda(\phi,\chi)$ it might be possible to circumvent the no-go theorem of 
Kaloper, Maddern and Olive for graceful exit \cite{kmo}
 in the context of Pre-Big Bang 
scenario. In a simple 
setting where we consider axion-dilaton and the potential due to the fluxes 
(maintaining S-duality invariance), we find that the no-go theorem still holds
(see the appendix for details).

Now we proceed to present an illustrative example for a cosmological solution. 
We compactify action (\ref{eq:eq1}) to $D=4$ on $T^6$ of constant moduli. 
Then set $B_{\mu\nu}^{(1)}=0,$ $B^{(1)}_{\mu\alpha}=0,$ 
$B_{\mu\alpha}^{(2)}=0$, $A_{\mu}^\alpha=0$ and the flux 
${\cal{H}}_{\alpha\beta\gamma}^{(1)}=0$. The resulting action is
\begin{eqnarray}
&&\widetilde{S}=\frac{1}{2}\int d^4x \sqrt{-g}\biggl[e^{-2\phi}\{R
+4(\partial \phi)^2\}-\biggl(\frac{1}{2}(\partial \chi)^2
+\frac{1}{2}\partial_\mu B_{\alpha\beta}^{(2)}\partial^\mu B^{(2)\alpha\beta}
\nonumber\\&&+\frac{1}{2}\partial_\mu \sigma^{(2)}\partial^\mu\sigma^{(2)}
+\frac{1}{12}{\cal{H}}_{\alpha\beta\gamma}^{(2)}
{\cal{H}}^{(2)\alpha\beta\gamma}\biggr)\biggr]\;.\label{eq:GV}
\end{eqnarray}
Where $H^{(2)}_{\mu\nu\lambda}$ has been dualized to yield $\sigma^{(2)}$. 
We have expressed the action (\ref{eq:GV}) in such a way that the two terms
in the curly bracket, multiplied by $e^{-2\phi}$, are the contributions from
the NS-NS states with 3-form set to zero. The rest of the terms come from the
R-R sector and these are not multiplied by over all factor of $e^{-2\phi}$.
Following Gasperini and Veneziano \cite{gvr},  we denote the second piece 
as $S_m$,  such that 
$\delta S_m/\delta\phi=0$. When we go to the cosmological setting all the 
fields depend on cosmic time $t$ and we can define $2\varphi=2\phi
-3{\rm{ln}}{\widetilde{a}}$, with 
$ds_4^2=-dt^2+{\widetilde{a}}^2(t)dx_i dx_j\delta_{ij}$. ${\widetilde{a}}(t)$ is the scale factor in this string frame metric. Thus scalars coming 
from R-R sector, including $\sigma_2$, do 
not couple to dilaton. Furthermore, if we define density and pressure as
\begin{equation}
T_0^0=\varrho\;,\ \ \ T_j^i=p\delta_j^i
\end{equation}
with $2{{\delta S_m}\over {\delta g^{\mu\nu}}}=T_{\mu\nu}$

the massless scalars are pressureless fluids. However, the flux contributes a 
negative pressure ($({\cal{H}}^{(2)})^2$). In order to establish correspondence
 with \cite{gvr}  we redefine the shifted dilaton $\bar{\phi}=2\varphi$ 
the relevant quantities are defined as follows: $\bar{\varrho}=\varrho a^3$, 
$\bar{p}=pa^3$. The resulting equations are 
\begin{eqnarray}
&&\dot{\bar{\phi}}^2-3{\widetilde{h}}^2=e^{\bar{\phi}}\bar{\varrho}\;,\\
&&\dot{{\widetilde{h}}}-{\widetilde{h}}\dot{\bar{\phi}}=\frac{1}{2}e^{\bar{\phi}}\bar{p}\;,\\
&&2\ddot{\bar{\phi}}-\dot{\bar{\phi}}^2-3{\widetilde{h}}^2=0
\end{eqnarray}
where ${{\widetilde{h}}}\equiv \frac{\dot{{\widetilde{a}}}}{{\widetilde{a}}}$.

The solution to shifted dilaton is
\begin{equation}
\bar{\phi}={\rm{ln}}\frac{12}{k}\frac{{\widetilde{a}}^{\sqrt{3}}}{(1-{\widetilde{a}}^{\sqrt{3}})^2}
\end{equation}
$k$ being the constant of integration. 
The scale factor is related to cosmic time through the following relation  
\begin{equation}
\frac{t}{t_0}=\frac{1}{\sqrt{3}}\bigl({\widetilde{a}}^{\sqrt{3}}-{\widetilde{a}}^{-\sqrt{3}}\bigr)
+2{\rm{ln}}{\widetilde{a}}
\end{equation}
and $t_0$ is another integration constant. The scaled density and pressure
(denoted by ${\bar \varrho}$ and $\bar p$ above) satisfy the relation
\begin{equation}
{\bar \varrho}=k{\widetilde{h}}^2 ~~~~~~{\rm and }~~~~{\bar p}=
-{2\over{\sqrt 3}}{{1-{\widetilde{a}}^{\sqrt{3}}}\over{1
+{\widetilde{a}}^{\sqrt 3}}}{\bar \varrho}\;.
\end{equation} 
 
We remind the reader that the equations of motion for R-R scalars lead to 
conservation law 
(corresponding charge conservation) like the previous case.  
We note that $\{{\widetilde{a}},\bar{\phi}\}$ satisfy scale factor duality 
during the entire 
epoch ${\widetilde{a}}(t)={\widetilde{a}}^{-1}(-t)$, 
$\bar{\varrho}({\widetilde{a}})=\bar{\varrho}({\widetilde{a}}^{-1})$ and 
$\bar{\phi}({\widetilde{a}})=\bar{\phi}({\widetilde{a}}^{-1})$. However as 
${\widetilde{a}}(t)\rightarrow 1$ dilaton assumes large value and 
it blows up at ${\widetilde{a}}=1$. 
Therefore, during this era we cannot trust the tree level string effective 
action and higher order corrections to the effective action 
have to be taken into account.
Furthermore, the truncated action Eq. (\ref{eq:GV}) is no longer manifestly 
S-duality 
invariant since we have removed some of the fields from the action associated 
with $SL(2,{\bf{R}})$ transformation. This is a scenario where S-duality is 
broken explicitly which specific choice of flux. It is not obvious at this 
stage whether one can address the graceful exit issue in this frame work if all
 the scalars appearing in action (\ref{eq:action6}) are included.
It requires exhaustive study of that effective action.

As we have discussed in the appendix, if we take another truncated version of 
the action, keeping different set of fields and the full potential, it is not 
possible to circumvent the no-go theorem. In view of the above discussions the 
present work offers new opportunities to address the graceful exit problem. 
Furthermore, other compactification schemes might be adopted to study cosmology
 from some of the points of view presented here.

\section{Summary and Discussion}
In this section we summarize our results and discuss some future directions of 
investigations. We have adopted toroidal compactification of the type IIB 
string effective action and included contribution of 3-form fluxes along 
compact directions. We have imposed a requirement that the reduced action 
remains S-duality invariant. Thus the axion-dilaton potential generated due to 
the presence of fluxes is $SL(2,{\bf{R}})$ invariant. Of course such a 
compactification scheme inherently contains tadpoles and supersymmetries are 
not preserved. We are aware of this aspect. However, our goal is to study the 
conventional string cosmology in this approach and address some of the issues. 
One of the interesting scenario, which is a special feature of 
graviton-axion-dilaton (string) cosmology is to understand accelerated expansion of the Universe according to Pre-Big Bang hypothesis. One of the 
issues encountered in the Pre-Big Bang approach is the graceful exit problem.
 In other words how does the Universe transit from its rapid accelerated expansion regime 
to the present FRW phase. It has been argued that such a transition is 
forbidden under certain circumstances due to no-go theorems. 
There were attempts to introduce 
phenomenological potentials to circumvent this difficulty. Moreover, it was 
very difficult to construct S-duality invariant axion-dilaton potential those 
days. However, with the present compactification scheme the potential is 
manifestly S-duality invariant. Therefore, we thought it to be appropriate to 
examine the graceful exit problem. We considered a truncated version of the 
reduced effective action (setting certain scalar fields to zero) where axion 
and dilaton which parameterize $\frac{SL(2,{\bf{R}})}{SO(2)}$ were retained in 
the action and the $SL(2,{\bf{R}})$ invariant potential was also retained. We 
find that the no-go theorem of Kaloper, Madden and Olive is still valid.

The four dimensional action contains the axion-dilaton potential. The action is
 that of a non-linear sigma model. It is possible that the global $
SL(2,{\bf{R}})$ symmetry might be spontaneously broken. Moreover, as alluded to
 in the introduction S-duality is not realized in Nature as an exact symmetry. 
However, it is believed that it is an exact symmetry of string theory. 
In this context, we may recall that when the M-theory is compactified on $T^2$
to type IIB theory in 9-dimensions, the resulting action can be cast in 
S-duality invariant form \cite{jhsb}. Moreover, "the coupling constants" (the
expectation values of dilaton and axion are the coupling constants) belong
to $T^2$ on which M-theory is compactified. Thus the two coupling constants
have geometric origin as has been argued by Schwarz \cite{jhsb}. 
Similarly, from the F-theory point of view axion-dilaton doublet of 
10-dimensional type IIB theory has also geometrical interpretations 
\cite{vafa}.

This 
symmetry must be broken below the string scale. There are several reasons why 
dilaton, which appears as a massless excitation along with the graviton, is not
 expected to remain massless at the present epoch. There are arguments from the 
cosmological stand point that dilaton must settle its ground state much before 
the era of nucleosynthesis. If we accept S-duality to be an exact stringy 
symmetry and that the dilaton does not remain massless at lower scales then the
 symmetry must be broken. Moreover, from the perspective of the standard model 
of particle physics axion is expected to be a light weakly interacting 
pseudoscalar particle. In a qualitative frame work string theory contains many 
axions. However, if we identify the S-duality partner of dilaton to be the 
axion that cosmologist search for and standard particle physics model 
introduces then it is important to understand how S-duality is broken.
If S-duality, realized non-linearly, is broken spontaneously then, in our 
model, it introduces a cosmological constant. We are unable to 
determine the symmetry breaking scale. We remind the reader that here one is 
discussing breaking of the non-compact $SL(2,{\bf{R}})$ symmetry. We are aware 
that in some cases one might encounter technical difficulties due to the 
noncompact nature of $SL(2,{\bf{R}})$. In the context of cosmological constant 
problem we would like to invoke the 
naturalness argument advanced by 't Hooft to argue that $\Lambda_c$ 
is small \cite{gerard}. 
Naturalness argument, to put qualitatively, says that if an exact 
symmetry dictates that a parameter in a theory is to vanish then that parameter
 will remain small when the symmetry is broken. In other words when a symmetry 
is restored if a parameter is forced to be zero the parameter will assume small
 value in the broken phase. It has been pointed out by 't Hooft that if 
cosmological constant, $\Lambda_c$, is set equal to zero in the Einstein-Hilbert 
action then there is no enhancement of any symmetry of the action. However, 
contrast this case with a theory which contains fermions. If fermion mass is 
set to zero, the action is invariant under chiral symmetry. In a more general 
setting we encounter situations such that exact symmetry require certain 
parameter have vanishing values. If some parameters assume small values, then the symmetries could be approximate.

We argue that cosmological constant arising out of spontaneous breaking of 
S-duality will remain small since in the unbroken phase it vanishes. However, 
our argument does not prevent from appearance of cosmological constant when
other symmetries are broken. At every stage of symmetry breaking there is
a contribution to the vacuum energy. The cosmological observations imply
presence of a vanishingly small cosmological constant. It is worth pursuing
our proposal to seek for a symmetry of the effective field theories that
describe phenomena today such that the naturalness argument may be invoked
to understand why $\Lambda _c$ is so small.  
In our study of cosmological scenarios, we have dealt with four dimensional 
effective actions. In such cases we arrived at the reduced action by 
compactifying on torus of constant modulus. Thus the volume modulus is set to 
constant by hand. In other words we have no way to stabilize the volume 
modulus.

It will be interesting the solve the Wheeler-De Witt equation for the 
cosmological case in the minisuperspace frame work. It is well known that the
graceful exit forbidden in the classical string cosmology, due to the no-go 
theorems, could be achieved
if one adopts the quantum version and solves the corresponding WDW equation
\cite{gmvq}. The axion-dilaton WDW equation has some novel features 
\cite{mmp,max} in that the S-duality symmetry may be exploited to obtain
the wave function for axion-dilaton purely from the group theoretic
consideration. In fact the dynamics of the two scalars in similar to 
motion of a particle on the surface of a pseudosphere \cite{max} in the
background FRW metric. We may hope that similar considerations could apply
when we include the presence of fluxes. However, it is obvious that the
cosmological version of eq.(34) cannot be solved when we consider the
corresponding WDW equation. Intuitively, we can argue that, although the 
interaction term is invariant under $SL(2,{\bf R})$ rotations, the 
axion-dilaton wave function of the corresponding  Hamiltonian cannot be
obtained as representations of $SL(2,{\bf R})$ as was the case in the 
absence of fluxes. If we denote the generators of $SL(2,{\bf R})$ as 
$\Sigma ^1=\pmatrix{1 & 0 \cr 0 & -1 \cr}$, $\Sigma ^2=\pmatrix{0 & 1 \cr 1 &
0 \cr }$ and $\Sigma ^3=\pmatrix{0 & 1 \cr -1 & 0 \cr}$, then the 
${\cal M}$-matrix can be expanded using unit matrix and $\Sigma ^i$ as a basis:
${\cal M}=v_0{\bf 1}+v_1\Sigma ^1+v_2\Sigma ^2+v_3\Sigma ^3$. It turns out
from structure of ${\cal M}$ is such that $v_3 =0$ and the three 
(time dependent)
coefficients satisfy the condition $v_0^2-v_1^2-v_2^2=1$, defining surface of
the pseudosphere as mentioned. The axion-dilaton kinetic energy term expressed 
in terms of $\{v_i;i=0,1,2\}$ can be written as the Laplace-Beltrami operator. 
When we go over to
the polar coordinates of the pseudosphere it becomes quite transparent that
this is the Casimir of $SL(2, {\bf R})$ \cite{max}. In the presence of the
fluxes, in general, the potential will not commute with any of the generators
of $SL(2, {\bf R})$ although it commutes with the Casimir. Therefore, if the
flux is along a special direction (say the term commutes with the compact
generator of $SL(2, {\bf R})$) then we can obtain wave functions (for the
"angular part") which are characterized by the Casimir and the eigenvalues of
the compact generator. These solutions will be infinite dimensional as is the
case with unitary representations of $SL(2,{\bf R})$. But it is not possible
to solve the full WDW equation in a closed form analytically for the scale 
factor part. However, it
is worthwhile to resort to some approximation like WKB and examine
various solutions to address the graceful exit problem. 

We have all along discussed $SL(2,{\bf{R}})$  duality symmetry in the context 
of string cosmology which has been our main focus. It is well known that, in 
string theory, the S-duality symmetry corresponds to the discrete 
$SL(2,{\bf{Z}})$ subgroup of $SL(2,{\bf{R}})$. The axion-dilaton potential will
 be further constrained under $SL(2,{\bf{Z}})$. It is argued that 
$SL(2,{\bf{Z}})$ is a robust symmetry of string theory. We have addressed the 
issue of S-duality symmetry breaking in this article. However, it is beyond the
 scope of present investigation to determine the scale of the symmetry 
breaking. In other words, although we argue that in the present epoch axion and
 dilaton are expected to acquire mass, we are unable to identify the scale of 
the symmetry breaking. We hope to take up some of the issues in future.

\bigskip

{\leftline{{\bf{Acknowledgements}}}} 

\smallskip
\smallskip

One of us (EK) would like to thank Professor Tohru Eguchi for valuable 
discussions and advices. JM would like to acknowledge fruitful discussions with
 Professors Ashok Das, Hikaru Kawai, Romesh Kaul, and P. K. Tripathy. The 
gracious hospitality of Professor Yoshihisa Kitazawa and the Theoretical High 
Energy Physics Division of KEK is gracefully acknowledged by JM.

 \section{APPENDIX: Graceful Exit Problem with Axion-Dilaton Potential }
 
 In this short appendix we discuss the graceful exit problem following Kaloper,
 Madden and Olive.\cite{kmo}  Note that our axion comes from R-R sector and 
therefore the
 relevant equations are modified appropriately. Our starting point is the
4-dimensional action eq. (34) written in terms of component fields in
string frame metric. We have pulled out an over all factor of $e^{-2\phi}$
and therefore the R-R field $\chi$ and the flux term get multiplied
accordingly.  The last term (potential is defined to be
$2\Lambda (\phi , \chi)$. The action is 
\begin{eqnarray}
S^{(4)}=\frac{1}{2}\int d^4x\sqrt{-g}e^{-2\phi}\bigg[R+4(\partial \phi)^2-
\frac{1}{2}e^{2\phi}(\partial \chi)^2 -
2\Lambda(\phi ,\chi) \bigg]\;.
\end{eqnarray}
In the cosmological case, all the fields are dependent on cosmic time 
(we consider the FRW metric with $k=0$). We denote the scale
factor in the string frame as
  $\widetilde{a}(t)$ and corresponding Hubble parameter as 
$\widetilde{h}=\frac{\dot{\widetilde{a}}}{\widetilde{a}}$, then the axion
equation of motion is 
\begin{equation}
\ddot{\chi}+3\widetilde{h}\dot{\chi}+2e^{-2\phi}\frac{\partial\Lambda(\phi,\chi)}{\partial \chi}=0\;.\label{eq:axion}
\end{equation}
The other equations are
\begin{eqnarray}
&&-3\widetilde{h}^2+6\dot{\phi}\widetilde{h}-2\dot{\phi}^2+
\frac{1}{4}e^{2\phi}\dot{\chi}^2+\Lambda (\phi , \chi)=0\;,\label{eq:EOM1}\\
&&4(\ddot{\phi}-\dot{\phi}^2+3\widetilde{h}\dot{\phi})-6\dot{\widetilde{h}}-12
\widetilde{h}^2+2\Lambda (\phi ,\chi)-
\frac{\partial \Lambda (\phi ,\chi)}{\partial \phi}=0\;,\label{eq:EOM2}\\
&&4(\ddot{\phi}-\dot{\phi}^2+2\widetilde{h}\dot{\phi})-4\dot{\widetilde{h}}-
6\widetilde{h}^2+2\Lambda (\phi ,\chi)-\frac{1}{2}
e^{2\phi}\dot{\chi}^2=0\;.\label{eq:EOM3}
\end{eqnarray}
Eq. (\ref{eq:EOM1}) is the Hamiltonian constraint which follows from variation 
of the lapse function, $N(t)$ and then setting $N(t)=1$ as is the standard 
practice. The other two equations are associated with variation of dilaton and 
the scale factor.

We recall that the Hamiltonian constraint is quadratic is $\dot{\phi}$ and it 
leads to 
\begin{equation}
\dot{\phi}=\frac{3\widetilde{h}\pm \sqrt{3\widetilde{h}^2+2\Lambda+\rho/2}}{2}
\label{eq:GV2}
\end{equation}
with $\rho=e^{2\phi}\dot{\chi}^2$. Moreover, as is well known the other two 
equations are utilized to eliminate $\ddot{\phi}$ and $\dot{\phi}^2$, which 
miraculously led to the Pre-Big Bang scenario, and we are left with an equation
 for $\dot{h}$
\begin{equation}
\dot{\widetilde{h}}=\pm\widetilde{h}\sqrt{3\widetilde{h}^2+2\Lambda+\rho/2}-\frac{1}{2}\frac{\partial\Lambda}{\partial \phi}+\frac{\rho}{4}\;.
\end{equation}

Where the sign is appropriately chosen from Eq. (\ref{eq:GV2}). The potential we recall is 
\begin{equation}
{\rm{Tr}}({\cal{H}}{\cal{H}}^T{\cal{M}})
\end{equation}
where ${\cal{H}}{\cal{H}}^T$ is to be understood as a $2\times2$ matrix with 
the (internal) tensor indices of ${\cal{H}}^{(1)}$ and ${\cal{H}}^{(2)}$ are 
appropriately  contracted. The case where the $SL(2,{\bf{R}})$ symmetry remains
 unbroken, the structure of the potential is such that it is positive 
everywhere and therefore, the no-go theorem of Kaloper, Madden and Olive 
remains valid.


\begin{thebibliography}{99}
\bibitem{book} M. B. Green, J. H. Schwarz and E. Witten, Superstring
Theory, Vol I and Vol II, Cambridge University Press, 1987.\\
 J. Polchinski, String Theory, Vol I and Vol II, Cambridge
University Press, 1998.\\
K. Becker, M. Becker and J. H. Schwarz, String Theory and M-Theory: A
Modern Introduction, Cambridge University Press, 2007. \\
B. Zwiebach, A First Course in String Theory, Cambridge University Press,
2004.
\bibitem{r1} A. Sen, Int. J. Mod. Phys. {\bf A 9}, 3707, (1994) 
[arXiv:hep-th/9402002].
\bibitem{r2} A. Giveon, M. Porrati and E. Rabinovici, Phys. Rep. {\bf C 244},
77 (1994) [arXiv:hep-th/9401139].
\bibitem{r3} E. Alvarez, L. Alvarez-Gaume and Y. Lozano, Nucl. Phys. Suppl.
{\bf 41}, 1  (1995) [arXiv:hep-th/9410237].
\bibitem{m1} J. Maharana and H. Singh, Phys. Lett. {\bf B 368}, 64 (1996) 
[arXiv:hep-th/9506213];
S. Kar, J. Maharana and H. Singh, Phys. Lett. {\bf B 374}, 43
(1996) [arXiv:hep-th/9507063].
\bibitem{m2} J. Maharana, Int. J. Mod. Phys. {\bf D 14}, 2245 (2005).
\bibitem{jim} J. E. Lidsey, On Cosmology and Symmetry of Dilaton-Axion 
Cosmology, [arXiv:gr-qc/9609063].
\bibitem{k1} E. Konishi, Axion-Dilaton Gauged S-Duality and Its Symmetry Breaking, arXiv:0710.1228 [hep-th].
\bibitem{ed} P. Svrcek and E. Witten, JHEP {\bf 0606}, 051 ,(2006), P. Svrcek,
Cosmological Constant and Axions in String Theory, [arXiv:hep-th/0607086].
\bibitem{kim} K. E. Kim, A Review on Axions and the Strong CP Problem,
arXiv:0909.2595 [hep-th].
\bibitem{k2}E. Konishi, Prog. Theor. Phys. {\bf{121}}, 1125 (2009) 
arXiv:0902.2565 [hep-th].
\bibitem{f1} K. Dasgupta, G. Rajesh and S. Sethi, JHEP {\bf 9908}, 023 (1999) 
[arXiv:hep-th/9908088].
\bibitem{f2} R. Bousso and J. Polchinski, JHEP, {\bf 0006}, 006 (2000) 
[arXiv:hep-th/0004134].
\bibitem{f3} S. B. Giddings, S. Kachru and J. Polchinski, Phys. Rev. 
{\bf D 66},
106006 (2002) [arXiv:hep-th/0105097].
\bibitem{frev1} M. R. Douglas and S. Kachru, Rev. Mod. Phys. {\bf 79}, 733
(2007) [arXiv:hep-th/0610102].
\bibitem{frev2} M. Grana, Phys. Rep. {\bf C 423}, 91 (2006) 
[arXiv:hep-th/0509003].
\bibitem{frev3} F. Denef, Constructing String Vacua (Les Houches Lecture),
arXiv:08003.1194 [hep-th].
\bibitem{kklt} S. Kachru, R. Kallosh, A. Linde and S. P. Trivedi, Phys. Rev.
{\bf D 68}, 046005 (2003) [arXiv:hep-th/0301240].
\bibitem{ss} J. Scherk and J. H. Schwarz, Nucl. Phys. {\bf B 153}, 61 (1979).
\bibitem{ms} J. Maharana and J. H. Schwarz, Nucl. Phys. {\bf B 390}, 3 (1993) 
[arXiv:hep-th/9207016].
\bibitem{hs} S. F. Hassan and A. Sen, Nucl. Phys. {\bf B 375}, 103 (1992) 
[arXiv:hep-th/9109038].
\bibitem{km1} N. Kaloper and R. C. Myers, JHEP {\bf 9905}, 010 (1999) 
[arXiv:hep-th/9901045].
\bibitem{jm1} J. Maharana, Phys. Lett. {\bf B 402}, 64 (1997) 
[arXiv:hep-th/9703009].
\bibitem{sources} P. K. Tripathy and S. P. Trivedi, JHEP {\bf 030}, 028 (2003) 
[arXiv:hep-th/0301139];
 S. Kachru, M. Schultz and S. P. Trivedi, JHEP {\bf 0310}, 007 (2003) 
[arXiv:hep-th/0201028];
J. Kumar and J. D. Wells, JHEP {\bf 0509}, 067 (2005) [arXiv:hep-th/0506252].
\bibitem{clw} E. J. Copeland, J. E. Lidsey and D. Wands, Phys. Rev. {\bf D 58},
043503 (1998) [arXiv:hep-th/9708153].
\bibitem{cr} E. J. Copeland, J. E. Lidsey and D. Wands, Phys. Rep. {\bf C 337},
343 (2000) [arXiv:hep-th/9909061].
\bibitem{gvr} M. Gasperini and G. Veneziano, Phys. Rep. {\bf C 373}, 1 (2003) 
[arXiv:hep-th/0207130].
\bibitem{gv1} M. Gasperini and G. Veneziano, Astropart. Phys. {\bf 1}, 317
(1993) [arXiv:hep-th/9211021].
\bibitem{bv} R. Brustein and G. Veneziano, Phys. Lett. {\bf B 329}, 429 (1994) 
[arXiv:hep-th/9403060];
N. Kaloper, R. Madden and K. A. Olive, Nucl. Phys. {\bf B 452}, 677 (1995) 
[arXiv:hep-th/9506027].
\bibitem{kmo} N. Kaloper, R. Madden and K. A. Olive, Phys. Lett. {\bf B 371},
34 (1996) [arXiv:hep-th/9510117].
\bibitem{v1} G. Veneziano, Phys. Lett. {\bf B 265}, 287(1991).
\bibitem{jm3} J. Maharana, Phys. Lett. {\bf B 372}, 53 (1996) 
[arXiv:hep-th/9511159]; J. Maharana
(unpublished work) 1996.
\bibitem{hull} C. M. Hull, Phys. Lett. {\bf B 357}, 545 (1995) 
[arXiv:hep-th/9506194].
\bibitem{jhs} J. H. Schwarz, Phys. Lett. {\bf B 360}, 13 (1995) 
[arXiv:hep-th/9508143].
\bibitem{ferr} S. Ferrara, C. Kounnas and M. Porrati, Phys. Lett. {\bf B 181},
263 (1986).
\bibitem{sroy}S. Roy, Int. J. Mod. Phys. {\bf{A 13}}, 4445 (1998) 
[arXiv:hep-th/9705016].
\bibitem{dp} T. Damour and A. M. Polyakov, Nucl. Phys. {\bf B 423}, 532 (1994) 
[arXiv:hep-th/9401069];
also see T. Damour and A. M. Polyakov, Gen. Rel. Grav. {\bf 26}, 1171 (1994) 
[arXiv:gr-qc/9411069].
\bibitem{mvc}K. Meissner and G. Veneziano, Mod. Phys. Lett. {\bf A 6}, 3397 
(1991); K. Meissner and G. Veneziano, Phys. Lett. {\bf B 267}, 33 (1991).
\bibitem{jhsb}J. H. Schwarz, Phys. Lett. {\bf B 367}, 97 (1996) 
[arXiv:hep-th/9510086].
\bibitem{vafa} C. Vafa, Nucl. Phys. {\bf B 469}, 403 (1996) 
[arXiv:hep-th/9602022].
\bibitem{gerard} G. 't Hooft, Under the Spell of the Gauge Principle, World
Scientific, 1994.
\bibitem{gmvq} M. Gasperini, J. Maharana and G. Veneziano, Nucl. Phys. 
{\bf B 472}, 349 (1996) [arXiv:hep-th/9602087].
\bibitem{mmp} J. Maharana, S. Mukherji and S. Panda, Mod. Phys. Lett. {\bf
A 12}, 447 (1996) [arXiv:hep-th/9701115].
\bibitem{max} J. Maharana, Int. J. Mod. Phys. {\bf A 20}, 1441 (2005) 
[arXiv:hep-th/0405039].

\end{thebibliography}
\end{document}